# *Viewpoint*
# Black Hole Evolution Traced Out with Loop Quantum Gravity

Carlo Rovelli

*CPT, CNRS, Aix-Marseille University, Toulon University, F-13288 Marseille, France*

**Abstract:** Loop Quantum gravity predicts that black holes evolve into white holes.
(Appeared in *Physics*, 11 (2018) 127, minimally edited.)

Black holes are remarkable entities. On the one hand, they have become familiar astrophysical objects. They are observed in large numbers via many channels: we have evidence of stellar-mass holes dancing around with a companion star, of gigantic holes at the center of galaxies pulling in spiraling disks of matter, and of black hole pairs merging emitting gravitational waves that have been detected. All of this is beautifully accounted for by Einstein's century-old general relativity. Yet, on the other hand, they remain highly mysterious: we see matter falling into them, and we are in the dark about what happens to this matter when it reaches the center of the hole.

Abhay Ashtekar and Javier Olmedo at Penn State University, and Parampreet Singh at Louisiana State University have taken an interesting step towards answering this question [1]. They have shown that loop quantum gravity—tentative theory of quantum gravity—predicts that spacetime continues across the center of the hole into a new region that has the geometry of the interior of a *white hole*, and is located in the future of the black hole (see Fig 1). A white hole is the time-reversed image of a black hole: in it, matter can only move outwards. The passage "across the center" into a future region is counterintuitive; it is possible thanks to the strong distortion of the spacetime geometry inside the hole allowed by general relativity.

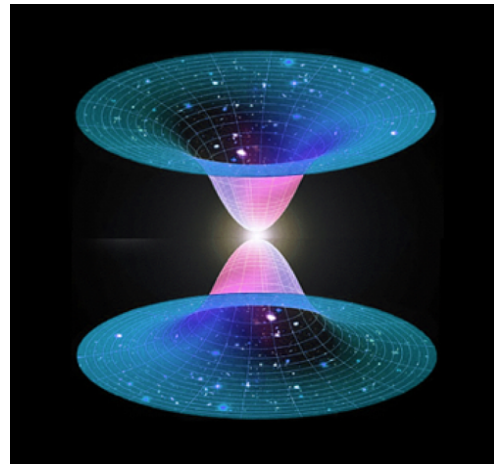

This result supports a scenario currently under investigation by several research groups: the future of all black holes may be to convert into a real *white* hole, from which the matter that fell inside can bounce out [2]. The fact that loop quantum gravity manages to show that the bounce happens is an indication that this theory has matured enough to tackle real-world situations.

***

*Figure 1: Artist rendering of black-to-white hole transition.*

The reason why we are in the dark about aspects of black hole physics is that quantum phenomena dominate at the center and in the future of these objects. Classical general relativity predicts that a black hole lives forever and its center is a "singularity" where space and time end. Both these predictions are un-realistic, because they disregard quantum effects. To correct them we need a quantum theory of gravity. There is no consensus on such a theory, but we have candidates, some of which are now reaching the point of allowing actual calculations on the quantum behavior of black holes. Loop quantum gravity, with its clean conceptual structure and well-defined mathematical formulation, is one such theory.

During the last years, a number of research groups have applied loop theory to explore the evolution of black holes [3-6]. These efforts have lead to a compelling picture, based on a black-to-white-hole transition scenario that can be summarized as follows [2]. At the center, space and time do not end into a singularity, but continue across a short transition region where the Einstein equations are violated by quantum effects. From this region, space and time emerge with the structure of a white hole interior, a possibility already suggested in the 1930s by John Lighton Synge [7]. As the hole's interior evolves, its external surface, or "horizon," slowly shrinks due to the emission of Hawking radiation—a phenomenon first described by Stephen Hawking. This shrinkage continues until the horizon reaches the Planck size (the characteristic scale of quantum gravity), or earlier [8,9], when a quantum transition ("quantum tunneling") happens at the horizon, turning it into the horizon of a white hole (see Fig 2). Thanks to the peculiar distorted relativistic geometry, the white hole interior born at the center is joined to the white horizon, completing the formation of the white hole. Remarkably, a part from the rapid quantum transition at the center and on the horizon, the rest of the spacetime geometry is an exact solution of the classical Einstein equations [9].

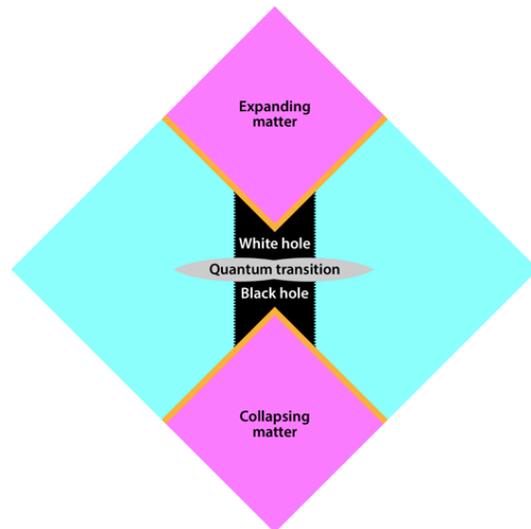

*Figure 2: The spacetime of a black hole turning into a white hole via a quantum transition..*

Loosely speaking, the full phenomenon is analogous to the bouncing of a ball. A ball falls to the ground, bounces, and then moves up. The motion after the bounce is the time-reversed version of the falling ball. Similarly, a black hole "bounces" and emerges as its time reversed version—a white hole. Collapsing matter does not disappear at the center: it bounces up through the white hole. Energy and information that fell into the black hole emerge from the white hole. The configuration where the compression is maximal, which separates the black hole from the white hole, is called a "Planck star" [8].

Because of the huge time distortion allowed by relativity, the time for the process to happen can be short (microseconds) measured from inside the hole, but long (billions of years) measured from the outside. Black holes might be bouncing stars seen in extreme slow motion.

∗∗∗

This is a compelling picture. It removes the singularity at a black hole's center and also resolves the paradox of the apparent disappearance of energy and information into a black hole. (For this, one must also realise that, contrary to a trendy assumption, the interior of a black hole can have many possible states—indistinguishable from the exterior as long as as a trapping horizon holds—even if its horizon is small [10].)

Ashtekar, Olmedo and Singh have shown that a crucial ingredient of this scenario, the transition at the center, follows from a genuine quantum gravity theory, namely loop theory. The basic idea that has lead to this result [3] is to use an approximation of the full loop-quantum-gravity equations similar to the one employed in previous work aimed at resolving the Big Bang singularity [11].

The Ashtekar-Olmedo-Singh model addresses only the transition at the center of the hole. To complete the picture, we also need the calculation of the tunneling at the horizon [2]. Preliminary steps in this direction have been taken [12,13], but the problem is open. Its solution would lead to a complete understanding of the quantum physics of black holes.

It is not implausible that empirical observations could support this scenario. Several observed astrophysical phenomena could be related to the black-to-white hole transition [14]. Among these are Fast Radio Bursts (FRBs) and certain high-energy cosmic rays [15]. Both could be produced by matter and photons that were trapped in black holes produced in the early universe and liberated by the black-to-white transition. For the moment, however, the astrophysical data are insufficient to determine whether the statistical properties of observed FRBs and cosmic rays confirm this hypothesis . Another intriguing possibility is that small holes produced by the black-to-white transition may be stable: in which case these "remnants" could be a component of dark matter [16].

We are only beginning to understand the quantum physics of black holes, but in this still speculative field the Ashtekar-Olmedo-Singh result gives us a welcome fixed point: loop gravity predicts that the interior of a black hole continues into a white hole.

The importance of any progress in this field goes beyond understanding black holes. The center of a black hole is where our current theory of spacetime, as of now given by Einstein's general relativity, fails. Understanding the physics of this region would mean understanding quantum space and quantum time.

[1] A. Ashtekar, J. Olmedo and P. Singh: Quantum Transfiguration of Kruskal Black Holes, Phys. Rev. Lett. 121, 241301 (2018); Quantum extension of the Kruskal spacetime, Phys. Rev. D 98, 126003 (2018).

[2] E. Bianchi, M. Christodoulou, F. D'Ambrosio, H. Haggard, C. Rovelli, White Holes as Remnants: A Surprising Scenario for the End of a Black Hole, Class. Quant. Grav. 35, 225003 (2018).

[3] 1. Modesto, L. Disappearance of the black hole singularity in loop quantum gravity. Phys. Rev. D - Part. Fields, Gravit. Cosmol. 70, 5 (2004).

[4] A. Ashtekar and M. Bojowald, Quantum Geometry and the Schwarzschild singularity, Class. Quant. Grav. 23, 391 (2006)

[5] M. Campiglia, R. Gambini and J. Pullin, Loop quantization of a spherically symmetric midi-superspaces: The interior problem, AIP Conf. Proc. 977, 52 (2008)

[6] A. Corichi and P. Singh, Loop quantum dynamics of Schwarzschild interior revisited, Class. Quant. Grav. 33, 055006 (2016)

[7] J. L. Synge. The Gravitational Field of a Particle. Proc. Irish Acad. A53, 83–114 (1950)

[8] C. Rovelli, V. Vidotto, Planck stars. Int. J. Mod. Phys. D23, 1442026 (2014).

[9] H. Haggard, C. Rovelli, Black hole fireworks: quantum-gravity effects outside the horizon spark black to white hole tunnelling, Physical Review D, 92.104020 (2015).

[10] C. Rovelli, Black holes have more states than those giving the Bekenstein-Hawking entropy: a simple argument", arXiv:1710.00218.

[11] I. Agullo, P. Singh, Loop Quantum Cosmology: A brief review, in *100 Years of General Relativity*, A. Ashtekar and J. Pullin eds., World Scientific 2018.

[12] M. Christodoulou, C. Rovelli, S. Speziale, I. Vilensky, Planck star tunneling time: An astrophysically relevant observable from background-free quantum gravity. Phys. Rev. D 94, 084035 (2016).

[13] M. Christodoulou, F. D'Ambrosio, Characteristic Time Scales for the Geometry Transition of a Black Hole to a White Hole from Spinfoams. (2018).

[14] A. Barrau, B. Bolliet, F. Vidotto, C. Weimer, Phenomenology of bouncing black holes in quantum gravity: A closer look. J. Cosmol. Astropart. Phys., 02, 022, (2016). A. Barrau, K. Martineau, F. Moulin, Status report on the phenomenology of black holes in loop quantum gravity: Evaporation, tunneling to white holes, dark matter and gravitational waves. Universe 4, 102 (2018).

[15] A. Barrau, C. Rovelli, F. Vidotto, F. Fast radio bursts and white hole signals. Phys. Rev. D 90, 127503 (2014).

[16] C. Rovelli, F. Vidotto, Small black/white hole stability and dark matter, Universe, 4 (2018).